# Data-driven Exploration of New Pressure-induced Superconductivity in PbBi$_2$Te$_4$ with Two Transition Temperatures

Ryo Matsumoto[a,d], Zhufeng Hou[b], Masanori Nagao[e], Shintaro Adachi[a], Hiroshi Hara[a,d], Hiromi Tanaka[f], Kazuki Nakamura[f], Ryo Murakami[f], Sayaka Yamamoto[f], Hiroyuki Takeya[a], Tetsuo Irifune[g], Kiyoyuki Terakura[c] and Yoshihiko Takano[a,d]

[a]*International Center for Materials Nanoarchitectonics (MANA),*
*National Institute for Materials Science, 1-2-1 Sengen, Tsukuba, Ibaraki 305-0047, Japan*
[b]*Research and Services Division of Materials Data and Integrated System (MaDIS),*
*National Institute for Materials Science, 1-2-1 Sengen, Tsukuba, Ibaraki 305-0047, Japan*
[c]*Center for Materials research by Information Integration (CMI$^2$),*
*National Institute for Materials Science, 1-2-1 Sengen, Tsukuba, Ibaraki 305-0047, Japan*
[d]*University of Tsukuba, 1-1-1 Tennodai, Tsukuba, Ibaraki 305-8577, Japan*
[e]*University of Yamanashi, 7-32 Miyamae, Kofu, Yamanashi 400-8511, Japan*
[f]*National Institute of Technology, Yonago College, 4448 Hikona, Yonago, Tottori 683-8502, Japan*
[g]*Geodynamics Research Center, Ehime University, Matsuyama, Ehime 790-8577, Japan*

**Abstract**

Candidates compounds for new thermoelectric and superconducting materials, which have narrow band gap and flat bands near band edges, were exhaustively searched by the high-throughput first-principles calculation from an inorganic materials database named AtomWork. We focused on $PbBi_2Te_4$ which has the similar electronic band structure and the same crystal structure with those of a pressure-induced superconductor $SnBi_2Se_4$ explored by the same data-driven approach. The $PbBi_2Te_4$ was successfully synthesized as single crystals using a melt and slow cooling method. The core level X-ray photoelectron spectroscopy analysis revealed $Pb^{2+}$, $Bi^{3+}$ and $Te^{2-}$ valence states in $PbBi_2Te_4$. The thermoelectric properties of the $PbBi_2Te_4$ sample were measured at ambient pressure and the electrical resistivity was also evaluated under high pressure using a diamond anvil cell with boron-doped diamond electrodes. The resistivity decreased with increase of the pressure, and two pressure-induced superconducting transitions were discovered at 3.4 K under 13.3 GPa and at 8.4 K under 21.7 GPa. The data-driven approach shows promising power to accelerate the discovery of new thermoelectric and superconducting materials.

# 1. Introduction

A data-driven approach based on high-throughput computation has recently been applied successfully to exploration of new functional materials such as battery materials, thermoelectric materials, superconductors, and so on. Once proper target quantities are selected, this approach may be more efficient than or at least complementary to traditional carpet-bombing type experiments based on experiences and inspirations of researchers [1-5]. We have reported a case-study of the data-driven approach thorough a discovery of pressure-induced superconductivity in a compound $SnBi_2Se_4$ selected by the high-throughput screening [6]. In this particular screening, the candidate compounds were explored according to a guideline that is characterized by specific band structures of "flat band" near the Fermi level, such as multivalley [7], pudding mold [8], and topological-type [9] structures. If such kinds of flat band are realized near the Fermi level, thermoelectric properties with high electrical conductivity and Seebeck coefficient would be enhanced [8,10]. If the flat band crosses the Fermi level, superconductivity would be realized due to high density of states (DOS) [11-13]. Single crystal of $SnBi_2Se_4$ actually synthesized showed an insulator to metal transition under 11 GPa, which shows good correspondence with the theoretical prediction. Moreover, a pressure-induced superconductivity was observed with maximum superconducting transition temperature ($T_c$) of 5.4 K under 63 GPa in accordance with our scenario. That work serves as a case study of the important first step for next-generation data-driven material science.

In the aforementioned data-driven approach, a high thermoelectric performance in $SnBi_2Se_4$ is expected under high pressure around its insulator to metal transition, since the band gap decreases by the applied pressure, and then the flat band approaches the Fermi level. If a certain compound with the same crystal structure and similar band shape has narrower band gap than that of $SnBi_2Se_4$, it will show superior thermoelectric property even at ambient pressure. Furthermore, it could be expected that superconductivity may appear at much lower pressure, compared with $SnBi_2Se_4$.

Based on these considerations, we focused on $PbBi_2Te_4$ as a target compound because it has same crystal structure and similar band structure with the band gap narrower than ~200 meV of $SnBi_2Se_4$. Superior thermoelectric properties at ambient pressure and the superconductivity under lower pressure could be expected in $PbBi_2Te_4$, compared with $SnBi_2Se_4$. In this study, we successfully synthesized the sample of $PbBi_2Te_4$ in a single crystal. The crystal structure, compositional ratio, and valence states of the $PbBi_2Te_4$ single crystal were analyzed by the powder X-ray diffraction (XRD), an energy dispersive X-ray spectrometry (EDX) and an X-ray photoelectron spectroscopy (XPS), respectively. The thermoelectric properties were measured at ambient pressure. The resistivity of the obtained sample was evaluated under high pressure using a diamond anvil cell (DAC) with boron-doped diamond electrodes [14-17].

## 2. Screening procedures in high-throughput first-principles calculations

1570 candidates were listed from the inorganic material database named AtomWork [18], based on the following restriction: abundant and nontoxic or less toxic constituent elements, and the number of atoms being less than 16 per primitive unit cell. The candidates were narrowed down by using the restriction of a narrow band gap and high DOS near the Fermi level. By this screening, the number of candidate compounds was reduced to 45. Finally, we checked whether the band gap decreases (or even the metallic behavior appears) or not under high pressure of 10 GPa, and screened out 27 promising compounds. Through the above screening procedures, $PbBi_2Te_4$ was chosen as a candidate for new thermoelectric and superconducting materials. The details of our screening scheme in the high-throughput first-principles calculations were given in our previous paper [6].

Figure 1 shows (a) the crystal structure of $PbBi_2Te_4$ with trigonal *R-3m* structure depicted by VESTA [19], (b) the band structure and (c) the total DOS of $PbBi_2Te_4$ obtained by the generalized gradient approximation with spin-orbit coupling. We can see that the band edges of $PbBi_2Te_4$ show flat shape near the Fermi level. This feature is quite similar with that of $SnBi_2Se_4$. Additionally, the band gap of 101 meV in $PbBi_2Te_4$ is less than half of 208 meV in $SnBi_2Se_4$. The feature in the band structure of $PbBi_2Te_4$ at ambient pressure is simular to that of $SnBi_2Se_4$ under high pressure of 5-10 GPa [6]. We could expect superior thermoelectric and superconducting properties under relatively low pressure for $PbBi_2Te_4$.

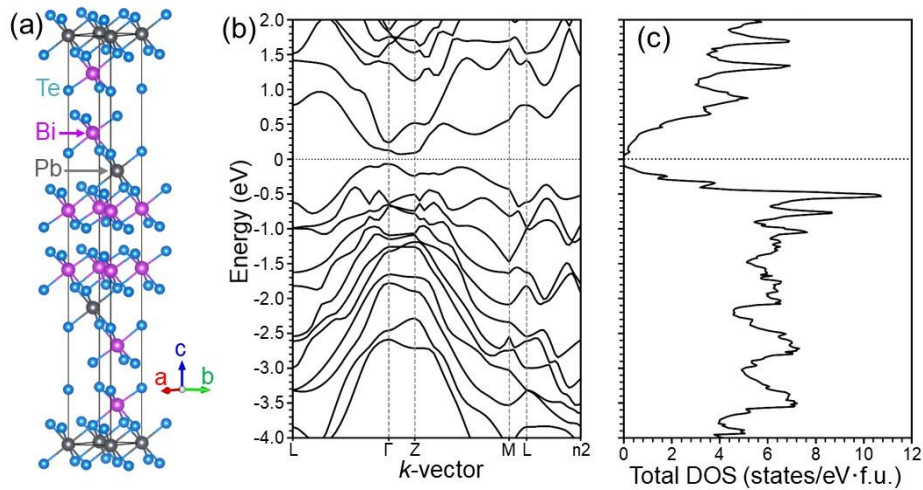

**Figure 1. (a) Crystal structure, (b) band structure and (c) total density of states (DOS) of $PbBi_2Te_4$ obtained by the generalized gradient approximation with spin-orbit coupling.**

## 3. Experimental procedures

### 3.1 Sample synthesis

Single crystals of $PbBi_2Te_4$ were grown by a melt and slow-cooling method. Starting materials of Pb grains, Bi grains, and Te chips were put into an evacuated quartz tube in the stoichiometric composition of $PbBi_2Te_4$. The ampoule was heated at 1000ºC for 1 hours, and then slowly cooled down to 800ºC for 20 hours. After keeping the temperature for 5 hours, the ampoule was cooled down to room temperature. The obtained samples were ground and loaded into an evacuated quartz tube again. The sample was heated at 500ºC for 50 hours for homogenization of $PbBi_2Te_4$ phase.

3.2 Characterization

The crystal structure of the obtained $PbBi_2Te_4$ samples was investigated by the powder XRD using the Mini Flex 600 (Rigaku) Cu K$\alpha$ radiation. The lattice constants were refined using the Conograph software [20]. The chemical composition of the sample was evaluated by an EDX analysis using the JSM-6010LA (JEOL). The valence state was estimated by the core level XPS analysis using AXIS-ULTRA DLD (Shimadzu/Kratos) with monochromatic Al K$\alpha$ X-ray radiation ($hv$ = 1486.6 eV), operating under a pressure of the order of $10^{-9}$ Torr. The samples were cleaved using scotch tape in a highly vacuumed pre-chamber in the order of $10^{-7}$ Torr. The analyzed area was approximately 1×1 mm$^2$. The binding energy scale was established by referencing the C 1$s$ value of adventitious carbon. The background signals were subtracted by using the active Shirley method implemented in COMPRO software [21]. The photoelectron peaks were analyzed by the pseudo-Voigt functions peak fitting.

3.3 Transport property measurement

Thermoelectric properties including the resistivity, Seebeck coefficient, and thermal conductivity were measured by a thermal transport option of physical property measurement system (PPMS/Quantum Design) under ambient pressure from 300 K to 2 K. The power factor and figure of merit were evaluated from the obtained parameters. Resistivity measurements of $PbBi_2Te_4$ single crystal under high pressure were performed using an originally designed DAC with boron-doped diamond electrodes [14-17]. Figure 2 shows an optical image of the sample space of our DAC. The sample was placed at the center of the bottom anvil where the boron-doped diamond electrodes were fabricated. The undoped diamond insulating layer covers the surface of the bottom anvil except for the sample space and electrical terminal. The details of the cell configuration were described in the literatures [16]. The cubic boron nitride powders with ruby manometer were used as a pressure-transmitting medium. The applied pressure values were estimated by the fluorescence from ruby powders [22] and the Raman spectrum from the culet of top diamond anvil [23] by an inVia Raman Microscope (RENISHAW).

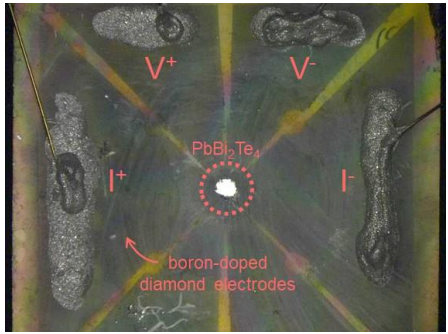

**Figure 2. Optical image of the sample space of DAC with boron-doped diamond electrodes**.

## 4. Results and discussion

4.1 Crystal structure, composition and valence state

Figure 3 shows a powder XRD pattern of the pulverized $PbBi_2Te_4$ single crystal. All observed peaks were well indexed to trigonal *R-3m* structure with lattice constants of $a = b = 4.42$ Å and $c = 41.57$ Å, without any impurity peaks. Here we note that if the sample is synthesized without the annealing process at 500ºC for 50 hours which is described in the experimental section, the $PbBi_4Te_7$ phase contaminated. The quantitative EDX analysis of the obtained single crystal showed the compositional ratio of $PbBi_{2.2}Te_{4.2}$, which indicates relatively excess Bi in the sample.

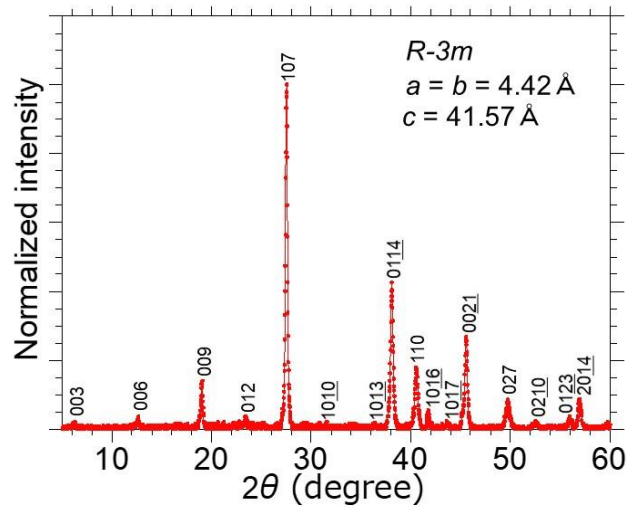

**Figure 3. Powder XRD pattern of pulverized $PbBi_2Te_4$ single crystal.**

The valence states of Pb, Bi and Te in $PbBi_2Te_4$ were investigated by XPS. Figure 4(a) shows a Pb 4f core-level spectrum of $PbBi_2Te_4$. There are two main peaks at 142.4 eV and 137.5 eV corresponding to Pb $4f_{5/2}$ and $4f_{7/2}$ with the valence state of $Pb^{2+}$ [24]. Figure 4(b) shows a Bi 4f core-level spectrum. The Bi 4f photoemission is split into two peaks, one at around 157.8 eV attributed to Bi $4f_{7/2}$ and the other at around 163.1 eV attributed to Bi $4f_{5/2}$ [25]. These main peak positions are corresponding to that of $Bi^{3+}$ valence state. The Te 3d region had two groups of peaks as shown in a fig.4(c), which were observed at 582.8 eV and 572.4 eV, indicating the existence of $Te^{2-}$, 586.5 eV and 576.1 eV of $Te^{4+}$ due to a surface oxide layer [26]. These $Pb^{2+}$, $Bi^{3+}$ and $Te^{2-}$ are consistent with the formal charge valence of $PbBi_2Te_4$. The Pb 4f and Bi 4f spectra contain small peaks at the higher binding energy region which may be due to the surface oxidization or the asymmetric feature of the main peaks [27]. If the asymmetricity is a Doniach-Sunjic line shape [28], it means the sample is metallic and has high DOS near the Fermi level [29]. Figure S1 in the supplemental materials shows the outcome of peak fitting using the Doniach-Sunjic line shaped pseudo-Voigt functions. The asymmetric parameter $α$ was 0.12 and 0.09 in Bi 4f and Pb 4f, respectively. Although the peak shape feature is consistent with our strategy of flat band model, further investigation is necessary to determine the reason of the asymmetricities.

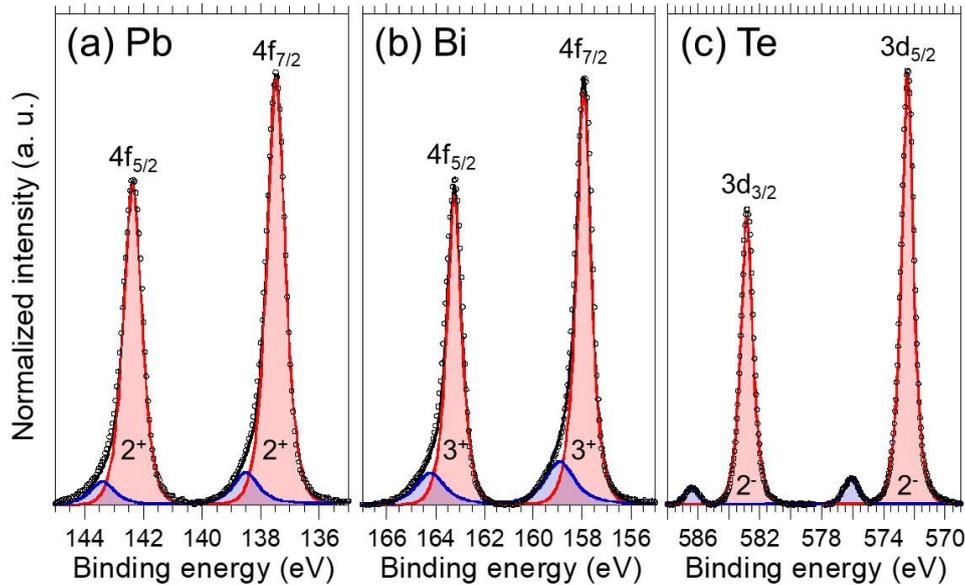

**Figure 4. High-resolution XPS spectra of (a) Bi 4f, (b)Pb 4f, and (c) Te 3d core levels in $PbBi_2Te_4$ single crystal.**

4.3 Thermoelectric properties

Figure 5 shows temperature dependence of the thermoelectric properties including (a) electrical resistivity, (b) Seebeck coefficient, (c) carrier concentration, and (d) thermal conductivity under ambient pressure for $PbBi_2Te_4$ (red plots), The results of $SnBi_2Se_4$ (blue plots) are also shown for comparison. $PbBi_2Te_4$ shows negative slope of resistivity toward lower temperature, namely metallic behavior. The absolute value of resistivity is much smaller than that of $SnBi_2Se_4$ [6]. The negative Seebeck coefficient and negative slope of the Hall voltage as a function of applied magnetic field indicate n-type nature of the sample, which may be caused by the excess Bi in the crystal. If the carrier type is tuned from n-type to p-type, the thermoelectric property could be enhanced because the valence band edge provides higher DOS near the Fermi level. A high pressure application, elemental substitution, or electric double layer transistor gating are effective for such kind of band tuning. Although the absolute value of the Seebeck coefficient in $PbBi_2Te_4$ is slightly smaller than that of SnBi$_2$Se$_4$, the thermal conductivity is almost same [6]. Consequently, the thermoelectric properties of the power factor and figure of merit were dramatically increased in PbBi$_2$Te$_4$ compared to SnBi$_2$Se$_4$. The highest values of the power factor ~ 100 μWm$^{-1}$K$^{-2}$ and figure of merit ~0.02 were obtained at 300 K. This tendency of high thermoelectric performance would be originated from the flat band with the small band gap, which is equivalent to the electronic state in SnBi$_2$Se$_4$ under high pressure [6].

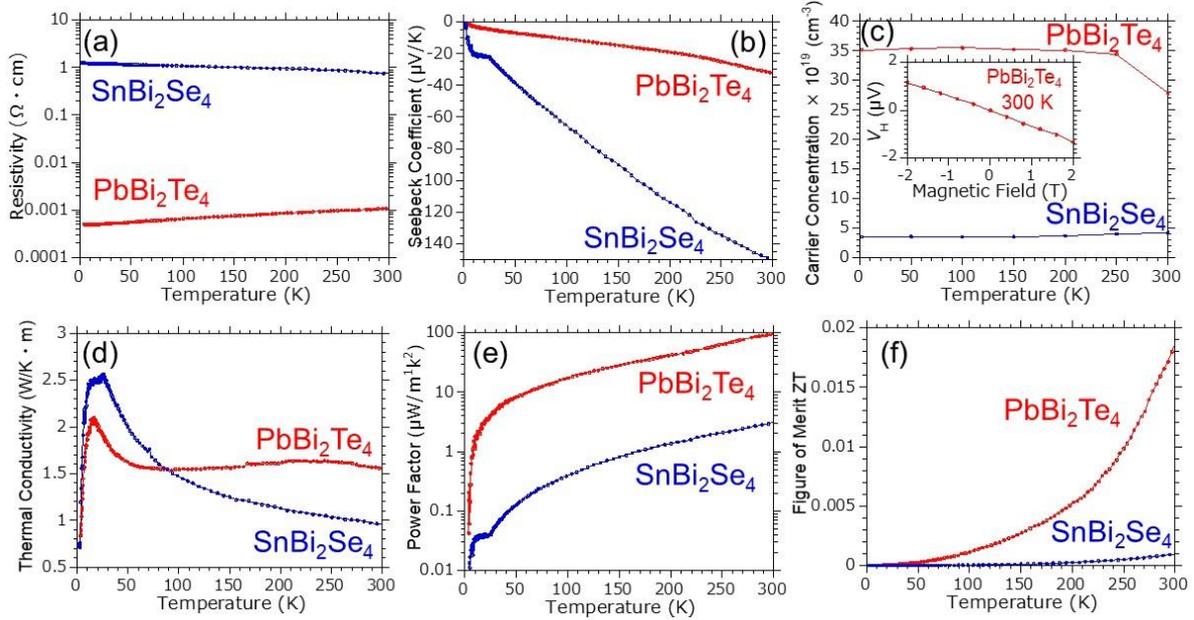

**Figure 5. Temperature dependence of thermoelectric properties in PbBi$_2$Te$_4$ and SnBi$_2$Se$_4$ under ambient pressure. (a) resistivity, (b) Seebeck coefficient, (c) carrier concentration (inset is a magnetic field dependence of Hall voltage at room temperature), (d) thermal conductivity, (e) power factor, and (f) figure of merit ZT.**

4.4 In-situ resistivity measurement under high pressure

Figure 6(a) shows a temperature dependence of resistivity for PbBi$_2$Te$_4$ under the various pressures from 1.0 GPa to 13.3 GPa. The sample has already exhibited metallic behavior under ambient pressure as shown in Fig. 5(a). The resistivity at 1.0 GPa shows also metallic behavior but with a small hump around 200 K. The resistivity and the intensity of hump decreased with the increase of the applied pressure. A pressure-induced superconductivity with clear zero resistivity was observed under 10.0 GPa. This critical pressure of the superconductivity is almost half of 20.2 GPa in SnBi$_2$Se$_4$. The maximum onset transition temperature ($T_c^{\mathrm{onset}}$) and zero-resistivity temperature ($T_c^{\mathrm{zero}}$) were 3.4 K and 2.4 K under 13.3 GPa, respectively.

The $T_c$ of PbBi$_2$Te$_4$ suddenly jumped up with the application of further pressure which implies the discovery of a second superconducting phase. A temperature dependence of resistivity from 13.3 GPa to 50.8 GPa is shown in a Fig. 6(b). The $T_c^{\mathrm{onset}}$ was enhanced from 3.4 K under 13.3

GPa to 8.1 K under 18.0 GPa. In the higher $T_c$ phase, the maximum $T_c^{onset}$ and $T_c^{zero}$ were 8.4 K and 7.9 K under 21.7 GPa, respectively. The temperature dependences of resistivity around the superconducting transitions were summarized in a Fig. 7. Indeed, the tendency of the two pressure-induced superconducting phases is quite similar to that of SnBi$_2$Se$_4$ [6]. Both the critical pressures for the higher and lower $T_c$ phases under ~10 GPa and ~20 GPa in PbBi$_2$Te$_4$ are almost half of ~20 GPa and ~40 GPa in the SnBi$_2$Se$_4$, respectively, due to the band gap difference. The higher $T_c$ in PbBi$_2$Te$_4$ would be originated from the higher DOS near the Fermi level because the higher pressure application decreases the DOS in SnBi$_2$Se$_4$ due to an increase of bandwidth.

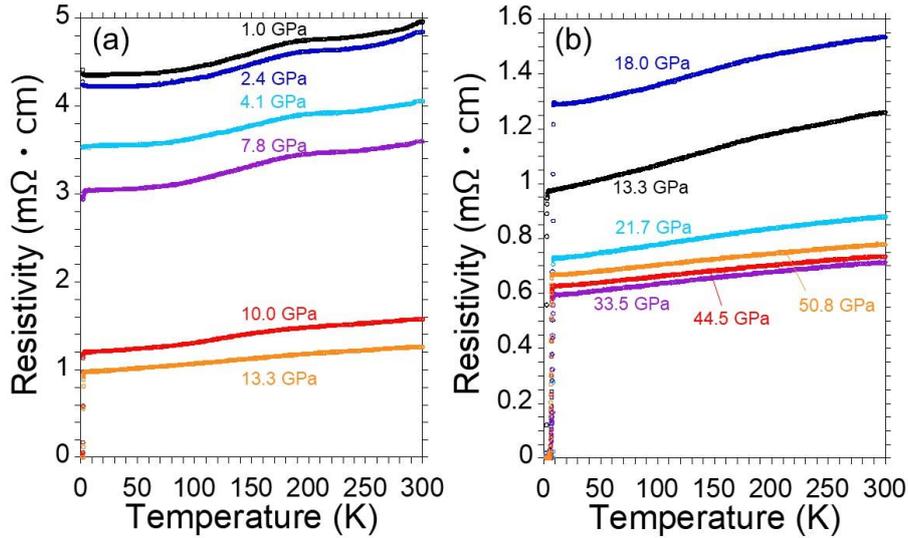

**Figure 6. Temperature dependence of resistivity in PbBi$_2$Te$_4$ under various pressures, (a) 1.0 - 13.3 GPa, (b) 13.3 - 50.8 GPa.**

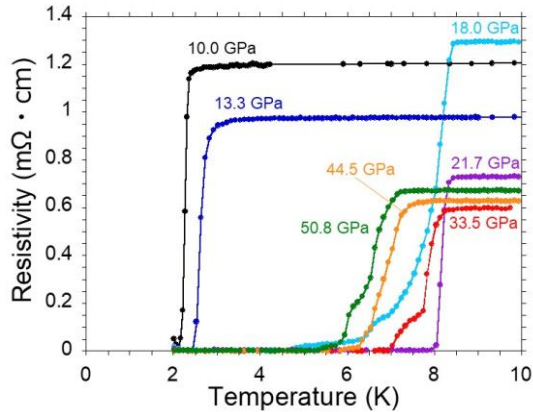

**Figure 7. Temperature dependence of resistivity around superconducting transitions in PbBi$_2$Te$_4$ under various pressures.**

Figure 8 shows temperature dependence of resistivity in various magnetic fields under (a) 13.3 GPa, (b) 21.7 GPa. Upper critical field $H_{c2}^{//ab}(0)$ values of lower and higher $T_c$ phases were estimated from the Werthamer-Helfand-Hohenberg (WHH) approximation [30] for the Type II superconductor in a dirty limit. A temperature dependence of $H_{c2}^{//ab}$ values were shown in the fig.8 (c). The $H_{c2}^{//ab}(0)$ were 2.4 T under 13.3 GPa (lower $T_c$ phase) and 5.9 T under 21.7 GPa (higher $T_c$ phase).

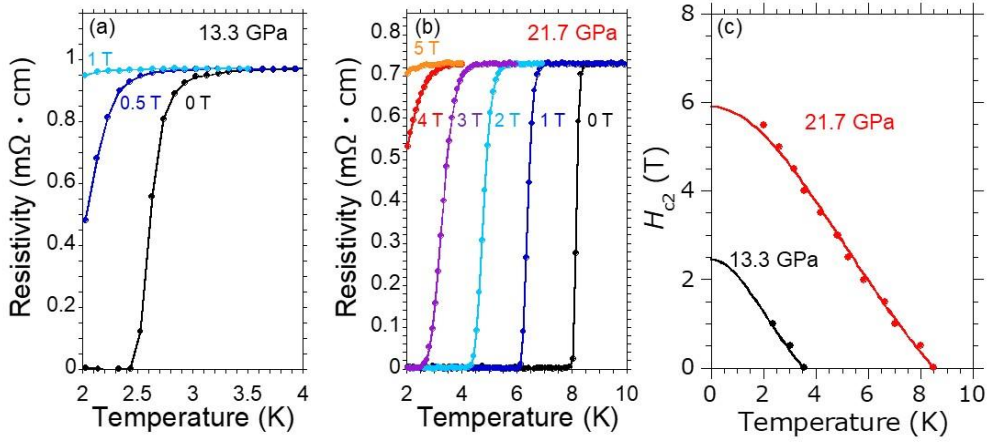

**Figure 8.** Temperature dependence of resistivity of PbBi$_2$Te$_4$ in magnetic fields under (a) 13.3 GPa, (b) 21.7 GPa. (c) temperature dependence of $H_{c2}^{//ab}$ values at 13.3 GPa and 21.7 GPa.

Figure 9 shows a pressure phase diagram of PbBi$_2$Te$_4$ single crystal. The resistivity of the sample is dramatically decreased by applying pressure. After that, the first superconducting phase was newly discovered under 10.0 GPa. In a higher pressure region above 18.0 GPa, we observed a higher $T_c$ phase. This superconductivity survived up to at least 50.8 GPa. The $T_c$ and $H_{c2}^{//ab}(0)$ values were almost independent with applied pressure.

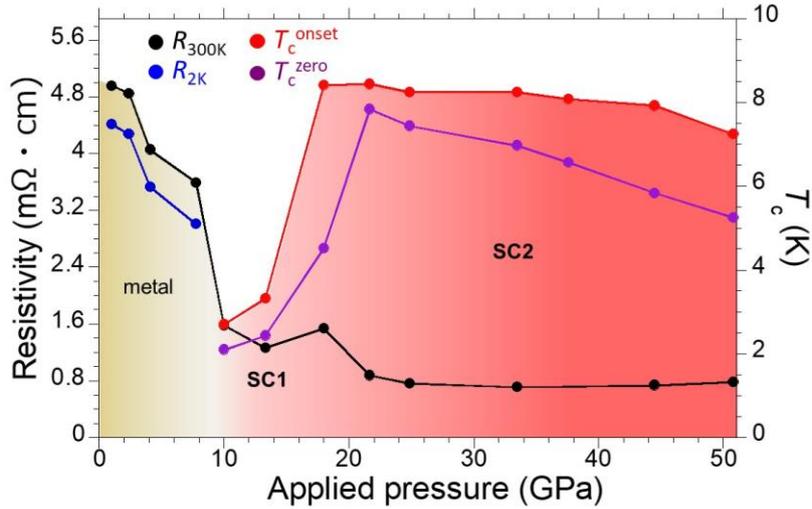

**Figure 9. Pressure phase diagram of PbBi$_2$Te$_4$.**

## 5. Conclusion

Among 27 compounds suggested by the data-driven approach, we focused on PbBi$_2$Te$_4$ from the viewpoint of band similarity to the pressure-induced superconductor SnBi$_2$Se$_4$ which was also chosen by the data-driven approach. The PbBi$_2$Te$_4$ has similar flat band feature with a smaller band gap, compared to that of SnBi$_2$Se$_4$. The synthesized PbBi$_2$Te$_4$ single crystal exhibited higher thermoelectric property at ambient pressure and higher $T_c$ value under high pressure. Especially, the superconductivity occurred under a pressure around 10 GPa lower than ~20 GPa in $SnBi_2Se_4$, in accordance with our scenario. The present work will serve as a case study of the important first-step for the next generation data-driven materials science.

**Acknowledgment**

This work was partly supported by JST CREST Grant No. JPMJCR16Q6, JST-Mirai Program Grant Number JPMJMI17A2, JSPS KAKENHI Grant Number JP17J05926, and the "Materials research by Information Integration" Initiative ($MI^2I$) project of the Support Program for Starting Up Innovation Hub from JST. A part of the fabrication process of diamond electrodes was supported by NIMS Nanofabrication Platform in Nanotechnology Platform Project sponsored by the Ministry of Education, Culture, Sports, Science and Technology (MEXT), Japan. The part of the high pressure experiments was supported by the Visiting Researcher's Program of Geodynamics Research Center, Ehime University. The computation in this study was performed on Numerical Materials Simulator at NIMS. The authors thank Dr. N. Kataoka (University of Okayama) for interpretation of XPS results.

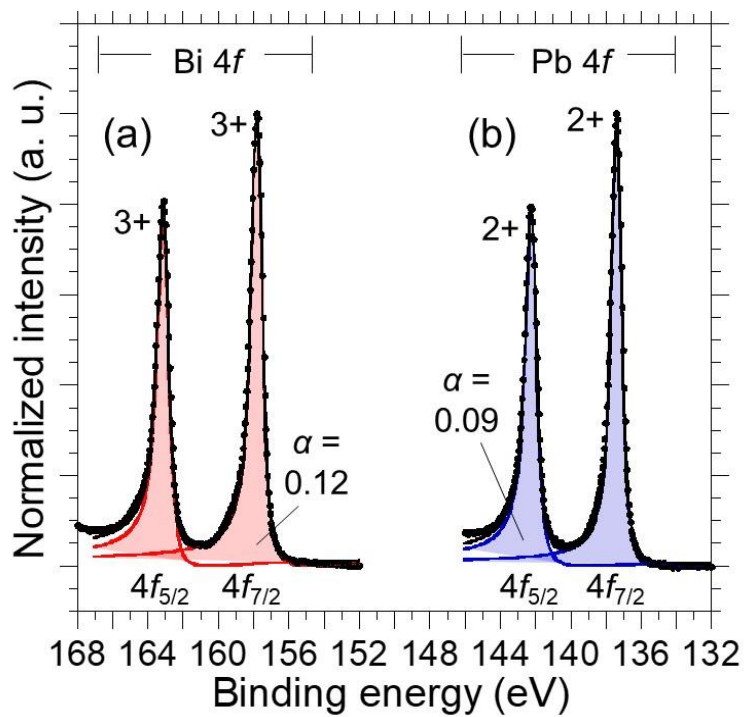

**Figure S1.** High-resolution XPS spectra of (a) Bi 4f, (b) Pb 4f with Doniach-Sunjic line shaped pseudo-Voigt functions.